\documentclass[USenglish,oneside,twocolumn,9pt]{article}

\usepackage{cite}
\usepackage{times}
\usepackage{amsmath,amssymb,amsfonts}
\usepackage{algorithmic}
\usepackage{graphicx}
\usepackage{textcomp}
\usepackage{xcolor}
\usepackage{listings}
\usepackage{upquote}
\usepackage{flushend}
\usepackage{pgfplots}
\usepackage{pgfplotstable}
\usepackage{filecontents}
\usepackage{booktabs}
\usepackage{multicol}
\usepackage[breaklinks,hidelinks]{hyperref}
\usepackage[nameinlink]{cleveref}
\usepackage{authblk}
\usepackage{enumitem}

\newcommand{\ignore}[1]{}
\def\name/{Fidelius} 

\newenvironment{denseitemize}{
\begin{itemize}[topsep=2pt, partopsep=0pt, leftmargin=1.5em]
  \setlength{\itemsep}{4pt} 
  \setlength{\parskip}{0pt}
  \setlength{\parsep}{0pt}
}{\end{itemize}}

\definecolor{editorGray}{rgb}{0.95, 0.95, 0.95}
\definecolor{editorOcher}{rgb}{1, 0.5, 0} 
\definecolor{editorGreen}{rgb}{0, 0.5, 0} 
\definecolor{editorRed}{rgb}{1, 0.1, 0.1}

\lstdefinelanguage{CSS}{
  keywords={color,background-image:,margin,padding,font,weight,display,position,top,left,right,bottom,list,style,border,size,white,space,min,width, transition:, transform:, transition-property, transition-duration, transition-timing-function},	
  sensitive=true,
  morecomment=[l]{//},
  morecomment=[s]{/*}{*/},
  morestring=[b]',
  morestring=[b]",
  alsoletter={:},
  alsodigit={-}
}

\lstdefinelanguage{JavaScript}{
  morekeywords={typeof, new, true, false, catch, function, return, null, catch, switch, var, if, in, while, do, else, case, break},
  morecomment=[s]{/*}{*/},
  morecomment=[l]//,
  morestring=[b]",
  morestring=[b]'
}

\lstdefinelanguage{HTML5}{
  language=html,
  sensitive=true,	
  alsoletter={<>=-},	
  morecomment=[s]{<!-}{-->},
  tag=[s],
  otherkeywords={
  >,
	<!DOCTYPE,
  </html, <html, <head, <title, </title, <style, </style, <link, </head, <meta, />,
	</body, <body,
	</div, <div, </div>, 
	</p, <p, </p>,
	</script, <script,
  <canvas, /canvas>, <svg, <rect, <animateTransform, </rect>, </svg>, <video, <source, <iframe, </iframe>, </video>, <image, </image>, <header, </header, <article, </article, <form, </form, <input, />
  },
  ndkeywords={secure=, sign=}
}

\lstdefinestyle{htmlcssjs} {%
  backgroundcolor=\color{editorGray},
  basicstyle={\footnotesize\ttfamily},   
  frame=b,
  xleftmargin={0.75cm},
  numbers=left,
  stepnumber=1,
  firstnumber=1,
  numberfirstline=true,	
  identifierstyle=\color{black},
  keywordstyle=\color{blue}\bfseries,
  ndkeywordstyle=\color{editorRed}\bfseries,
  commentstyle=\color{brown}\ttfamily,
  language=HTML5,
  alsolanguage=JavaScript,
  alsodigit={.:;},	
  tabsize=2,
  showtabs=false,
  showspaces=false,
  showstringspaces=false,
  extendedchars=true,
  breaklines=true,
  literate=%
  {Ö}{{\"O}}1
  {Ä}{{\"A}}1
  {Ü}{{\"U}}1
  {ß}{{\ss}}1
  {ü}{{\"u}}1
  {ä}{{\"a}}1
  {ö}{{\"o}}1
}

\def\BibTeX{{\rm B\kern-.05em{\sc i\kern-.025em b}\kern-.08em
    T\kern-.1667em\lower.7ex\hbox{E}\kern-.125emX}}
\begin{document}

\title{\textbf{\name/: Protecting User Secrets \\ from Compromised Browsers }}

\author[1]{\small Saba Eskandarian}
\author[1]{Jonathan Cogan}
\author[1]{Sawyer Birnbaum}
\author[1]{Peh Chang Wei Brandon}
\author[1]{\authorcr Dillon Franke}
\author[1]{Forest Fraser}
\author[1]{Gaspar Garcia}
\author[1]{Eric Gong}
\author[1]{Hung T. Nguyen}
\author[1]{\authorcr Taresh K. Sethi}
\author[1]{Vishal Subbiah}
\affil[1]{Stanford University}
\author[2]{Michael Backes}
\affil[2]{CISPA Helmholtz Center for Information Security}
\author[1,2]{Giancarlo Pellegrino}
\author[1]{Dan Boneh}

\maketitle

\begin{abstract}
Users regularly enter sensitive data, such as passwords, credit card numbers, or tax information, into the browser window.  While modern browsers provide powerful client-side privacy measures to protect this data, none of these defenses prevent a browser compromised by malware from stealing it. In this work, we present \name/, a new architecture that uses trusted hardware enclaves integrated into the browser to enable protection of user secrets during web browsing sessions, even if the \emph{entire underlying browser and OS} are fully controlled by a malicious attacker.

\name/ solves many challenges involved in providing protection for browsers in a fully malicious environment, offering support for integrity and privacy for form data, JavaScript execution, XMLHttpRequests, and protected web storage, while minimizing the TCB. Moreover, interactions between the enclave and the browser, the keyboard, and the display all require new protocols, each with their own security considerations. Finally, \name/ takes into account UI considerations to ensure a consistent and simple interface for both developers and users. 

As part of this project, we develop the first open source system that provides a trusted path from input and output peripherals to a hardware enclave with no reliance on additional hypervisor security assumptions.  These components may be of independent interest and useful to future projects. 

We implement and evaluate \name/ to measure its performance overhead, finding that \name/ imposes acceptable overhead on page load and user interaction for secured pages and has \emph{no impact} on pages and page components that do not use its enhanced security features. 
\end{abstract}

\section{Introduction}

The web has long been plagued by malware that
infects end-user machines with the explicit goal of stealing sensitive
data that users enter into their browser window.  
Some recent examples
include TrickBot and Vega Stealer, which are man-in-the-browser malware
designed to steal banking credentials and credit card numbers.  
Generally speaking, once malware infects the user's machine, it can effectively
steal all user data entered into the browser.  Modern browsers have
responded with a variety of defenses aimed at ensuring browser integrity.
However, once the machine is compromised, there is little
that the browser can do to protect user data from a key logger. 

In this paper we present a practical architecture, called \name/, that
helps web sites ensure that user data entered into the browser cannot
be stolen by end-user malware, no matter how deeply the malware is
embedded into the system.  
When using \name/, users can safely enter data into the browser
without fear of it being stolen by malware, provided that the hardware
enclave we use satisfies the security requirements.

Hardware enclaves, such as Intel's SGX, have recently been used to
provide security for a variety of applications, even in case of
compromise~\cite{haven,panoply,sgxkernel,scone,graphene,seabed,securekeeper,iron,prochlo,enclavedb,zerotrace,ryoan,azure-confidential,opaque,oblix}.
An enclave provides an execution environment that is isolated from the
rest of the system (more on this below).  Moreover, the enclave can
attest its code to a remote web site.

One could imagine running an entire browser in an enclave to isolate it
from OS-level malware, but this would be a poor design --
any browser vulnerability would lead to malware inside the enclave,
which would completely compromise the design.

\subsection{Our Contributions}

\name/ contains three components, discussed in detail in the following
sections:
(1) a small trusted functionality running inside an isolated hardware enclave, 
(2) a trusted path to I/O devices like the keyboard and the display, and
(3) a small browser component that interacts with the hardware enclave.

A trusted path from the hardware enclave to I/O devices is essential
for a system like \name/.  First, this is needed to prevent an
OS-level malware from intercepting the data on its way to and from the
I/O device.  More importantly, the system must prevent out-of-enclave
malware from displaying UI elements that fool the user into entering
sensitive data where the malware can read it.  Beyond protecting web
input fields, the system must protect the entire web form to ensure
that the malware does not, for example, swap the ``username'' and
``password'' labels and cause the user to enter her password into the
username field.

We implement a prototype trusted path to the keyboard using a
Raspberry Pi Zero that sits between the user's machine and the
keyboard and implements a secure channel between the keyboard
and the hardware enclave.  We implement a trusted path to the
display using a Raspberry Pi~3 that sits between the graphics card and
the display.  The Raspberry Pi~3 overlays a trusted image from the
hardware enclave on top of the standard HDMI video sent to the display
from the graphics card.  We discuss details in Section~\ref{imppath}.
Our trusted path system is open source
and available for other projects to use. 
We note that we can not use SGXIO~\cite{sgxio}, 
an SGX trusted I/O project, 
because that system uses hypervisors, which
may be compromised in our threat model.

Another complication is the need to run client-side JavaScript on
sensitive form fields.  For example, a web site may use client-side
JavaScript to ensure that a credit card checksum is valid, and alert
the user if not.  Similarly, many sites use client-side JavaScript to
display a password strength meter. \name/ should not prevent these
scripts from performing as intended.  Several projects have already
explored running a JavaScript interpreter in a hardware enclave.
Examples include TrustJS~\cite{trustjs} and
Secureworker~\cite{secureworker}.  Our work uses the ability to run
JavaScript in an enclave as a building block to enable privacy for
user inputs in web applications.  The challenge is to do so while
keeping the trusted enclave -- the TCB -- small.

To address all these challenges, this paper makes the following contributions:
\begin{denseitemize}
\item The design of \name/, a system for protecting user secrets entered 
  into a browser in a fully-compromised environment.
\item A simple interface for web developers to 
  enable \name/'s security features.
\item The first open design and implementation of a trusted path enabling
  a hardware enclave to interact with I/O devices such as 
  a display and a keyboard from a fully compromised machine. 
\item A browser component that enables a hardware enclave to interact 
with protected DOM elements while keeping the enclave
component small.
\item An open-source implementation and evaluation of \name/ for
practical use cases.
\end{denseitemize}

\ignore{
\medskip\noindent{\bf Related work.}  The problem of safely entering
data into an end-user application has been studied extensively, and we
survey the related work in Section~\ref{sec:related}.  We note that
Microsoft's 2003 NGSCB~\cite{NGSCB} architecture, also known as
Palladium, was a commercial attempt at a solution.  However, none of
the proposed systems were deployed widely, despite the importance
of the problem.  We believe that the new hardware capabilities used in
\name/ make it a practical and deployable solution that can be
integrated into existing web sites with relatively little effort.

\medskip\noindent{\bf Security of hardware enclaves.}
We built \name/ using the hardware enclave provided by Intel's SGX.
SGX has recently come under several side-channel
attacks~\cite{sgxpectre,foreshadow}, making the current implementation
of SGX insufficiently secure for \name/.  However, Intel is updating
SGX using firmware and hardware updates with the goal of preventing
these side-channel attacks.  In time, it is likely that SGX can be
made sufficiently secure to satisfy the requirements needed for
\name/.  Even if not, other enclave architectures are available, such
as Sanctum for RISC-V~\cite{sanctum} or possibly a separate
co-processor for security operations.
}

\section{Background} 

A \emph{hardware enclave} provides developers with the abstraction of a secure portion of the processor that can verifiably run a trusted code base (TCB) and protect its limited memory from a malicious or compromised OS \cite{CD16, SGXRef}. The hardware handles the process of entering and exiting an enclave and hiding the activity of the enclave while non-enclave code runs. Enclave code invariably requires access to OS resources such as networking and user or file I/O, so developers specify an interface between the enclave and the OS. In SGX, the platform we use for our implementation, the functions made available by this interface are called \textit{OCALLs} and \textit{ECALLs}. OCALLs are made from inside the enclave to the untrusted application, usually for procedures requiring resources managed by the OS, such as file access or output to a display. ECALLs allow code outside the TCB to call the enclave to execute trusted code.

An enclave proves that it runs an untampered version of the desired code through a \textit{remote attestation} mechanism. Attestation loosely involves an enclave providing a signed hash of its initial state (including the running code), which a server compares with the expected value and rejects if there is any evidence of a corrupted program. In order to persist data to disk when an enclave closes or crashes, SGX also provides a data \emph{sealing} functionality that encrypts and authenticates the data for later recovery by a new instance of the enclave.

Finally, one of the key features of enclaves is the protection of memory. An enclave gives developers a small memory region inaccessible to the OS and only available when execution enters the enclave. In this memory, the trusted code can keep secrets from an untrusted OS that otherwise controls the machine. SGX provides approximately 90MB of protected memory. Unfortunately, a number of side-channel attacks have been shown to break the abstraction of fully-protected enclave memory. We briefly discuss these attacks and accompanying defenses below and in \Cref{sec:related}. 

\noindent{\bf Security of hardware enclaves.}
We built \name/ using the hardware enclave provided by Intel's SGX.
SGX has recently come under several side-channel
attacks~\cite{sgxpectre,foreshadow}, making the current implementation
of SGX insufficiently secure for \name/.  However, Intel is updating
SGX using firmware and hardware updates with the goal of preventing
these side-channel attacks.  In time, it is likely that SGX can be
made sufficiently secure to satisfy the requirements needed for
\name/.  Even if not, other enclave architectures are available, such
as Sanctum for RISC-V~\cite{sanctum} or possibly a separate
co-processor for security operations.

\section{Threat Model}\label{threatmodel}

We leverage a trusted hardware enclave to protect against a network attacker who additionally has full control of the operating system (OS) on the computer running \name/. We assume that our attacker has the power to examine and modify unprotected memory, communication with peripherals/network devices, and communication between the trusted and untrusted components of the system. Moreover, it can maliciously interrupt the execution of an enclave. Note that an OS-level attacker can always launch an indefinite denial of service attack against an enclave, but such an attack does not compromise privacy. 

We assume that the I/O devices used with the computer are not compromised and that the dongles we add to keyboards/displays follow the behavior we describe. We could assume that there is a trusted initial setup phase where the devices can exchange keys and other setup parameters with the enclave. This corresponds to a setting where a user buys a new computer, sets it up with the necessary peripherals, and then connects to the internet, at which point the machine immediately falls victim to malware. Alternatively, this honest setup assumption could easily be avoided with an attestation/key exchange step between the peripherals and the enclave. We discuss both options in Section~\ref{setupio}.

\noindent \textbf{Overview of Security Goals}. 
We would like to provide the security guarantee that any user data entered via a trusted input will never be visible to an attacker, and, except in the case of denial of service, the data received by the server will correspond to that sent by the user, e.g. it will not be modified, shuffled, etc. Moreover, the enclave will only send data to an authenticated server, and a server will only send data to a legitimate enclave. Finally, we wish for all the low-level protocols of our system to be protected against tampering, replay, and other attacks launched by the compromised OS. 

The remote server in our setting cooperates to secure the user by providing correct web application code to be run in the enclave. We are primarily concerned with the security of user secrets locally on a compromised device, but this does include ensuring that secrets are not sent out to an attacker. 

\noindent \textbf{Overview of Usability Goals}. Although our work is merely a prototype of \name/, we intend for it to be fully functional and to defend not only against technical attacks on security but also against user interface tricks aiming to mislead a user into divulging secrets to a malicious party. This task looms particularly important in our mixed setting where trusted input/output come through the same channels as their untrusted counterparts. In particular, we must make sure a user knows whether the input they are typing is protected or not, what data the remote server expects to receive, and where the private data will eventually be sent. We leave the task of optimizing the user experience to future work, but also aim to provide a tool which can be used ``as-is.''

We also want to provide a usable interface for developers that deviates only minimally from standard web development practices. As such, we endeavor to add only the minimal extensions or limitations to current web design techniques to support our security requirements. 

\noindent \textbf{Enumeration of Attacks}. After describing the system in detail in subsequent sections, we discuss why \name/ satisfies our security goals. Here we briefly list the different classes of non-trivial attacks against which we plan to defend. Refer to \Cref{security} for details on the attacks and how we defend against them. 

\begin{denseitemize}
\item[-] \emph{Enclave omission attack}: The attacker fakes use of an enclave.
\item[-] \emph{Enclave misuse attack}: The attacker abuses Enclave ECALLs for unexpected behavior.
\item[-] \emph{Page tampering attack}: The attacker modifies protected page elements or JavaScript.
\item[-] \emph{Redirection attack}: The attacker fakes the origin to which trusted data is sent.
\item[-] \emph{Storage tampering attack}: The attacker reads, modifies, deletes, or rolls back persistent storage.
\item[-] \emph{Mode switching attack}: The attacker makes unauthorized entry/exits from private keyboard mode.
\item[-] \emph{Replay attack}: The attacker replays private key presses or display overlays.
\item[-] \emph{Input manipulation attack}: The attacker forges or manipulates placement of protected input fields.
\item[-] \emph{Timing attack}: The attacker gains side-channel information from the timing of display updates or keyboard events.
\end{denseitemize}

\noindent \textbf{Security Non-Goals}. \name/ provides the tools necessary to form the basis of a secure web application, focusing on protecting user inputs and computation over them. We do not provide a full framework for secure web applications or a generic tool for protecting existing web applications. In particular, we do not protect against developers who decide to run insecure, leaky, or malicious JavaScript code inside an enclave, but we do provide a simple developer interface to protect security-critical components of applications.

We assume the security of the trusted hardware platform and that the enclave hides the contents of its protected memory pages and CPU registers from an attacker with control of the OS, so side channel attacks on the enclave~\cite{sgxpectre,foreshadow} are also out of the scope of this work. We discuss side channel attacks and mitigations for SGX in \Cref{sec:related}. Physical attackers who tamper with the internal functionality of our devices also lie outside our threat model, but we note that our trusted devices seem to be robust against \emph{opportunistic} physical attackers that do not tamper with hardware internals but can, for example, attach a usb keylogger to a computer. The SGX hardware itself is also designed to resist advanced hardware attackers. 

Finally, we do not address how the honest server protects sensitive data once the user's inputs reach it. Our goal is to protect data from compromise on the client side or in transit to the server. Once safely delivered to the correct origin, other measures must be taken to protect user data. For example, we do not defend against a server who receives secrets from the user and then displays them in untrusted HTML sent back to the browser.

\section{Architecture Overview}
\begin{figure}
\centering
\includegraphics[width=0.95\linewidth]{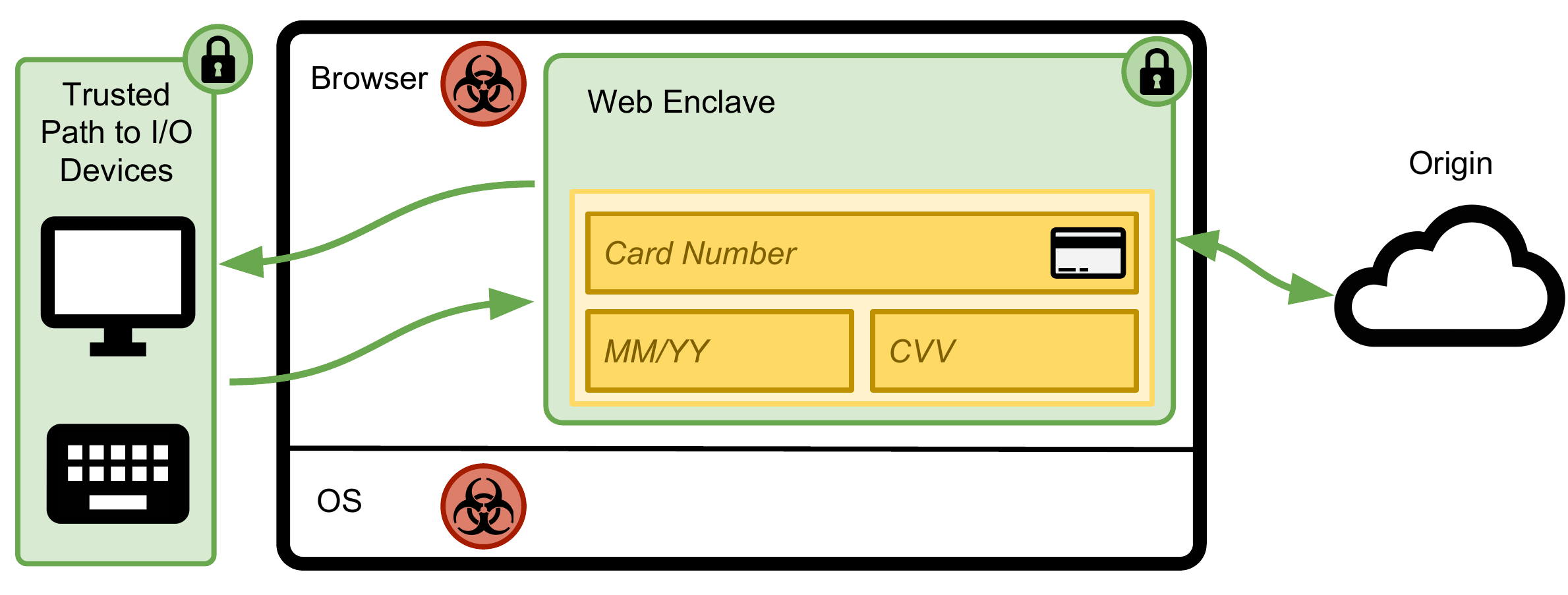}
\caption{\small Overview of \name/. The web enclave, embedded in a malicious browser and OS, communicates with the user through our trusted I/O path and securely sends data to a remote origin. We assume that both the web browser and the OS are compromised.}
\label{fig:arch}
\end{figure}

The goal of \name/ is to establish a trusted path between a user and the remote server behind a web application. To achieve this goal, \name/ relies on two core components: a trusted user I/O path and a \emph{web enclave}. In practice, this involve subsystems for a secure keyboard, a secure video display, a browser component to interact with a hardware enclave, and the enclave itself. Figure~\ref{fig:arch} gives an overview of the components of \name/. 

\subsection{Trusted User I/O Path} 

The trusted user I/O path consists of a keyboard and display with a trusted dongle placed between them and the computer running \name/. Each device consists of trusted and untrusted modes. The untrusted modes operate exactly the same as in an unmodified system. The trusted keyboard mode, when activated, sends a constant stream of encrypted keystrokes to the enclave. The enclave decrypts and updates the state of the relevant trusted input field. The trusted and untrusted display modes are active in parallel, and the trusted mode consists of a series of overlays sent encrypted from the enclave to the display. Overlays include rendered DOM subtrees (including, if any, the protected user inputs) placed over the untrusted display output as well as a dedicated portion of the screen inaccessible to untrusted content. We cover these functionalities and details of the protocols used to secure them in \Cref{sec:trustedpath}. Finally, both trusted devices have LEDs that notify the user when a trusted path is established and ready to collect user input. Our system relies, in part, on users not typing secrets on the keyboard when these security indicator lights are off.  This ensures that only the enclave has access to secrets entered on the keyboard.  We note, however, that several works have studied the effectiveness of security indicators in directing user behavior~\cite{WI05,emperor} and found that users often ignore them. We briefly discuss potential alternatives in Section~\ref{discussion}, but leave the orthogonal problem of designing a better user interface -- one that is more difficult to ignore -- to future work. 

\subsection{Web Enclave}
A web enclave is essentially a hardware enclave running a minimalistic, trusted browser engine bound to a single web origin. A browser using a web enclave delegates the management and rendering of portions of a DOM tree and the execution of client-side scripts, e.g. JavaScript and Web Assembly, to the enclave. In addition, the web enclave can send and receive encrypted messages to and from trusted devices and the origin server. Finally, the web enclave provides client-side script APIs to access the DOM subtree, secure storage, and secure HTTP communication.

When a user loads a web page, \name/ checks whether the page contains HTML tags that need to be protected, e.g., secure HTML forms. If it does, it initiates a web enclave, runs remote attestation between that enclave and the server, and validates the identity of the server. Once this process completes, \name/ loads the HTML tags it needs to protect into the web enclave and verifies their signatures. Then, when the user accesses a protected tag, e.g. with a mouse click, \name/ gives control to the enclave, which in turn activates the devices' trusted mode. The trusted mode LEDs are turned on, informing the user that the trusted path is ready to securely collect user input.

Web enclaves provide two main ways to send protected messages to a remote server: directly through an encrypted form submission or programmatically via an XMLHttpRequest API. When a user clicks a form's submit button, the web browser notifies the enclave of this event. Then, the web enclave encodes the form data following HTML form norms\footnote{See, \url{https://www.w3.org/TR/html5/sec-forms.html}}, encrypts that data, and signs it. The encrypted form is passed to the web browser, which sends it to the remote server. When a script needs to send messages to the server, it can use the XMLHttpRequest web API. The web enclave XMLHttpRequest API interface is similar to that implemented by web browsers; however, it  encrypts sensitive fields such as the request body and custom HTTP headers. HTTP responses are sent by the server in encrypted form. The enclave will automatically decrypt responses and resume execution of the JavaScript function waiting for the response.

\section{Interface Design}\label{UIsection}

This section describes the interfaces that \name/ provides for end-users and developers who wish to consume or create protected web applications. Here we describe only how \name/ appears to users and developers, deferring technical details of how it works to subsequent sections. 
\subsection{User Interface}
\begin{figure}
\centering
\includegraphics[width=0.9\linewidth]{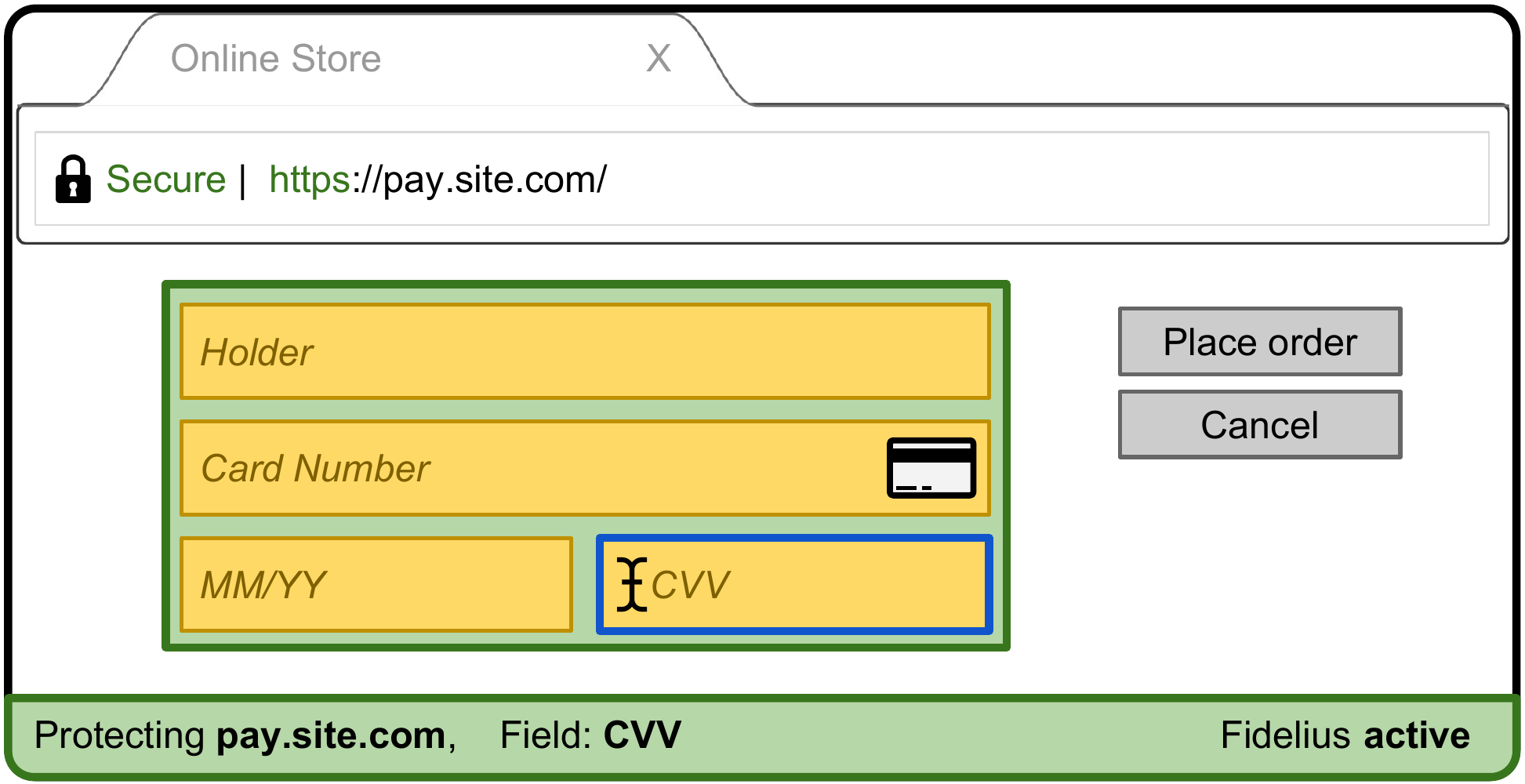}
\caption{\small Design of \name/'s user interface. The green area is the trusted display overlay.}
\label{UIpicture}
\end{figure}

The primary challenge in designing an interface for a system with a mix of trusted and untrusted components lies in distinguishing the trusted parts from the untrusted parts in a way that cannot be faked by an attacker. Our solution is to dedicate a small part of the screen to the web enclave, rendering that portion of the screen inaccessible to the OS while the trusted display is active, as indicated by an LED outside the display. Outside of this region, user interaction with \name/ does not differ at all from interactions with a typical web application. \Cref{UIpicture} shows the design of \name/'s user interface in use on a sample payment page. Trusted input fields do not have any special visual features that distinguish them from other inputs. Instead, the dedicated trusted region of the screen displays information that defends against attacks which make use of UI manipulation to fool a user into giving sensitive data to an attacker.

There are two important pieces of information shown in the protected display region. First, we must ensure that the user sends sensitive information only to the intended destination and avoids attacks like changing the contents of the url bar or picture-in-picture attacks~\cite{picinpic}. We achieve this by including the origin of the web enclave in the trusted region. In \Cref{UIpicture}, the trusted region shows that the web enclave is connected to \texttt{pay.site.com}.

Second, we must ensure that users can distinguish real trusted inputs from untrusted ones and that an attacker cannot fool the user by changing the untrusted text surrounding a trusted input field. This could include attacks where untrusted input fields are made to look just like trusted ones (which in fact is the case by default in \name/) or, for example, where the username and password prompts before two inputs are switched, causing the user's password to be processed as a username, which potentially receives far less protection after being sent to the server. We protect against this class of attacks by displaying a name for each trusted input in the dedicated display region when that field has focus. 
This serves to indicate to the user that the current input field is trusted. It also protects against any attack involving shuffling of input field labels to fool a user or cause incorrect data to be sent to the server because the descriptive name for each input field lies outside the reach of an attacker. 

\subsection{Developer Interface}
Design of a developer interface must provide an easy to use and backwards compatible way for developers to access the features of \name/. Our developer interface requires \emph{no changes} for pages or components of pages that do not make use of \name/'s features. Developers who wish to provide stronger security guarantees to \name/ users include additional attributes in existing HTML tags directing \name/ to use the web enclave in rendering and interacting with the content of those tags. \Cref{codePicture} shows an example of an HTML page supporting \name/. 

\begin{lstlisting}[style=htmlcssjs,caption={\small \name/-enabled code for the online payment web page. In red, the new HTML attributes required by \name/.}, captionpos=b, label={codePicture}]
<html>
 <head> [...] </head>
 <body> 
  <form action="submit_data"
        name="payment"
        method="POST"
        secure="True" sign="tX5ReRzE42Qw">
    <input type="text" 
           value="Holder" name="holder" />
    <input type="text" 
           value="Card Number" name="card"/>
    <input type="text" 
           value="MM/YY" name="exp"/>
    <input type="text" 
           value="CVV" name="cvv"/>
  </form>
  <div class="btn"><p>Place order</p></div> 
  <div class="btn"><p>Cancel</p></div>
  <script type="text/JavaScript" 
          src="validator.js" 
          secure="True" sign="Fi3Rt9mq2ff0">
  </script>
 </body>
</html>
\end{lstlisting}

\name/ currently supports \texttt{<form>}, \texttt{<input>}, and \texttt{<script>} tags. To mark any of these tags as compatible with \name/, developers add a \texttt{secure} attribute to the tag. In the case of \texttt{<script>} and \texttt{<form>} tags, a signature over the content of the tag is included in a \texttt{sign} attribute, to be verified with respect to the server's public key inside the enclave as described in \Cref{features}. The signature ensures that the form and script contents have not been modified by malware before they were passed to the enclave. The signature is not needed for \texttt{<input>} tags because the signature on a form includes the inputs contained within it. \texttt{<input>} tags also require a \texttt{name} attribute to be shown in the trusted component of the display when that input has focus. 

JavaScript included in secure \texttt{<script>} tags runs on an interpreter inside the web enclave with different scope than untrusted code running in the browser. Trusted JavaScript has access to its own memory and its own web APIs for secure storage and secure HTTP requests, but it cannot directly access the memory or web APIs available to untrusted JavaScript. Trusted and untrusted JavaScript can, however, make calls to each other and pass information between each other as needed using an interface similar to the \texttt{postMessage} cross-origin message passing API. 

\name/ enforces a strict same-origin policy for web enclaves, so network communication originating or ending in an enclave can only come from its specified origin. By default, the origin of HTML tags is inherited from the web page. In general, the origin is derived from the initial URL of the page. However, for tags such as \texttt{<form>} and \texttt{<script>}, the origin is derived from the \texttt{action} and \texttt{src} attributes respecively. The origin specified here is not authenticated and therefore susceptible to tampering. We discuss the process by which a web enclave connects to remote servers and verifies their legitimacy in \Cref{features}.

\section{Trusted Path for User I/O}\label{sec:trustedpath}
In this section, we describe the building blocks to create and manage a trusted path connecting a keyboard, display, and web enclave. Specifically, we cover device setup, communication patterns between devices, and the structure of individual messages passed between devices. 

Although we develop our trusted I/O path in the context of the larger \name/ system and focus our discussion on web applications, it is important to note that the trusted path is fundamentally a separate system from the web enclave. In other words, although the two systems interact closely in the design of \name/, the trusted path has applications outside the web and can be run on its own as well. To our knowledge, this is the first system to provide a trusted path to the user for both input and output relying only on assumptions about enclave security. We cover the details of how we realize the trusted peripherals in hardware dongles in \Cref{sec:imp}.

\subsection{Setup}\label{setupio}
In order to securely communicate, the web enclave and peripherals (or the dongles connected to them) must have a shared key. One option is to operate in a threat model with an initial trusted phase where we assume the computer is not yet compromised. Pre-shared keys are exchanged when the user configures the computer for the first time. Devices store the key in an internal memory, and the enclave seals the shared keys for future retrieval. The key can be accessed only by the enclave directly and not by user-provided JavaScript running inside it.

In the more realistic setting where new peripherals can be introduced to a computer over time, we must protect against attacks that involve introduction of malicious periphal devices. In this setting, we need \name/-compatible devices to include a trusted component that can perform an attestation with the enclave to prove its legitimacy before exchanging keys. Note that this attestation must occur in both directions -- from enclave to keyboard and from keyboard to enclave -- or the device that does not attest can be faked by an attacker. 

\subsection{Trusted Communication}
The process of switching between trusted and untrusted modes presents an interesting security challenge. An authentication procedure between the enclave and the trusted devices can ensure that only the enclave initiates switches between trusted and untrusted modes, but this ignores the larger problem that the enclave must rely on the untrusted OS to inform it when an event has happened that necessitates switching modes. Avoiding that necessity would require moving a prohibitively large fraction of the browser and UI into an enclave. Our solution has two parts and relies on making the user aware of when key presses produce trusted or untrusted input. First, we include a light on each dongle that turns on only when the keyboard or display are in trusted mode. This alone, however, does not suffice to solve the problem, as an attacker could mount a ``rapid switching'' attack where it jumps in and out of trusted mode faster than the user can perceive or react, leading to parts of the user's input being leaked by untrusted input. Even worse, rapid switching between modes may occur quickly enough to not be noticable to a user monitoring the lights. To prevent this attack, we force a short delay when switching out of trusted mode. This ensures the user will have time to notice and react when a switch occurs. 

The enclave switches devices in and out of trusted mode by sending one of two reserved messages which are simply fixed strings that they interpret as commands to change the trust setting. When in trusted mode, messages between the enclave and the peripherals are encrypted as described in \Cref{sec:msgstructure}.

Since the timing of key presses can reveal sensitive information about what keys are being pressed~\cite{keytiming}, we must also avoid leaking timing information while in trusted input mode. We do this by having the keyboard send a constant stream of key presses where most contain only an encryption of a dummy value that indicates no key pressed. As long as the fixed frequency of key presses exceeds the pace at which a user types, the user experience is unaffected by this protection. Since user key presses typically result in changes on the display, we update the display contents at the same rate as we read keyboard inputs. 

Our trusted input design in many ways mirrors that of Bumpy~\cite{bumpy} and SGX-USB~\cite{sgxusb}, which also provide generic trusted user input using similar techniques but do not provide the web functionality that we do. In contrast to our work, Bumpy does not display any trusted user input. SGX-USB allows for generic I/O but does not solve the problem of mixing trusted and untrusted content in a user interface as we do in both our keyboard and display. Neither system has source code available. We improve on the features of both works by protecting against timing attacks on encrypted data sent from trusted input devices. 

\subsection{Message structure}\label{sec:msgstructure}
Messages sent in the trusted communication protocol described above must include safeguards against replay attacks. To do this, we include a counter in every message sent, so that the same count never repeats twice. Counters are maintained on a per-device and per-origin basis, so every message between the enclave and the keyboard or display must include a counter value and the name of the origin in addition to the encrypted key press or overlay itself. 

\section{Web Enclave}\label{features}

In this section we cover the details of the web enclave. First, we provide an overview of the state transitions of a web enclave. Next, we present the protocols for remote attestation, origin authentication, and exchange of key material. Finally, we present the details of the operations: secure HTML forms, JavaScript code execution, secure network communication, and persistent storage across web enclave executions. 

\subsection{Web Enclave State Machine}

\begin{figure}
    \centering
    \includegraphics[width=0.9\linewidth]{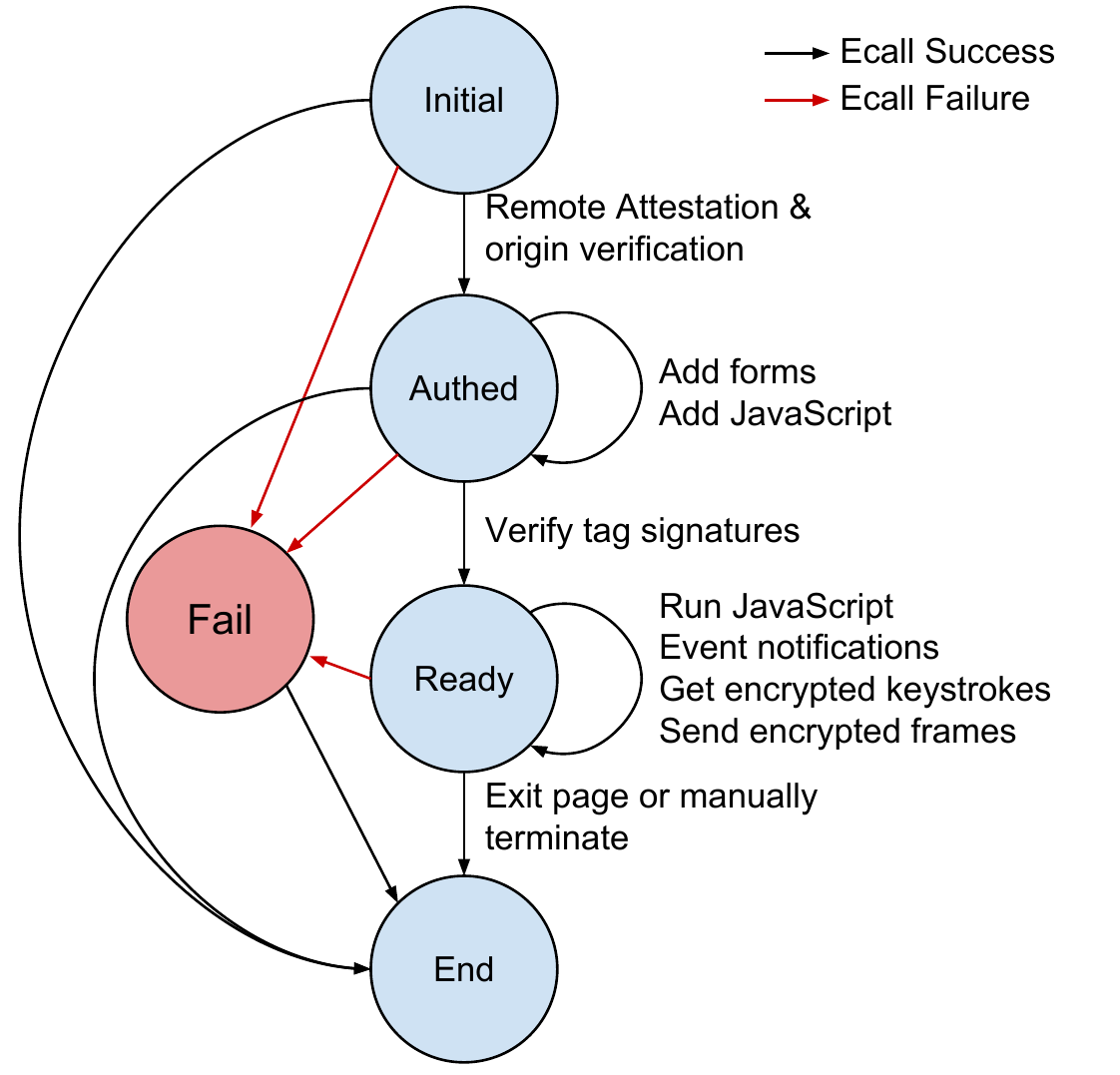}
    \caption{\small State machine representing web enclave behavior.}
    \label{fig:webenc_fsm}
\end{figure}

The web enclave implements the state machine in \Cref{fig:webenc_fsm}. At any point, it can be in one of the following five states: \emph{initial}, \emph{authenticated}, \emph{ready}, \emph{end}, and \emph{fail}. Transitions are caused by ECALLs. Each state has a list of accepted ECALLs. For example, the initial state accepts only ECALLs for the remote attestation and origin validation. Other ECALLs bring the web enclave to the fail state. No other transition is possible from this state, and the enclave needs to be terminated after reaching it. 

\name/ creates a web enclave when it finds any \texttt{<form>} or \texttt{<script>} tags with the \texttt{secure} attribute set. Then it derives the origin of the tags that need to be protected. By default, the origin of the tags are inherited from the web page they belong to, i.e., the domain and port of the URL. However, for tags such as \texttt{<form>} and \texttt{<script>}, the origin is derived from the \texttt{action} and \texttt{src} attributes respecively. Tags can have different origins. While it is possible to create one web enclave for each origin, the current version of the web enclave assumes that all protected components on a page communicate with the same origin.

Once the origin has been determined, \name/ passes the origin to the web enclave and performs remote attestation and origin validation, after which the enclave and the origin can share a symmetric key. This key will be used to encrypt any communication between the enclave and the origin, so any network manipulation or monitoring will only result in an attacker recovering encrypted data for which it does not have the key. As a result, the rest of the network stack can remain outside the enclave in untrusted code. In order to verify an origin, the enclave must have the corresponding public key, either as a hard-coded value or, more realistically, by verifying a certificate signed by a hard-coded authority. 

At this point, the web enclave is in the \emph{authenticated} state. \name/ retrieves the tags with the \texttt{secure} attribute set and loads them into the enclave. These operations do not cause a state transition. The only ECALL that causes a valid transition from this state is verification of the signatures. If the validation of all signatures succeeds, the enclave enters the \emph{ready} state. From this point on, the enclave is fully operational and can decrypt keyboard inputs, prepare encrypted outputs for the display, execute JavaScript functions, and take/release control of the trusted path upon focus and blur events respectively. 

\subsection{Features}
Once an enclave has successfully entered the ready state, the full functionality of \name/ becomes available to the web application. \name/ supports secure HTML forms, JavaScript execution, secure network communication, and persistent protected storage. 

\noindent\textbf{Secure HTML Forms.}
When parsing a page, \name/ finds \texttt{<form>} tags with the secure attribute and, after verifying the provided signature using the server's public key, creates a \texttt{form} data structure inside the enclave to keep track of the form and each of the inputs inside it. We currently store server public keys inside the enclave but could replace this with root certificates instead. When the user highlights an input inside a given form, the browser notifies the enclave. The enclave switches the keyboard from untrusted to trusted input mode (see \Cref{sec:trustedpath} for details), and subsequent user key presses modify the state of the highlighted input field. As mentioned in \Cref{UIsection}, various defenses at the interface level protect against attacks that an attacker could mount by modifying untrusted content between the enclave and the user. By pushing these defensess into the UI, we allow ourselves to keep many components of the browser outside of the enclave and dramatically reduce \name/'s TCB. For example, monitoring of mouse movements and placement of forms on the page can be managed outside the enclave, and tampering/dishonesty with these elements will be detected by a user who notices the inconsistency between what she sees on the screen and the content of the trusted overlay. 

Submission of HTML forms involves encrypting the content of the form as one blob using the shared key negotiated during attestation and sending that to the server. 

\noindent\textbf{Javascript.}
We run a JavaScript interpreter inside the enclave but leave out heavy components like the event loop. When a trusted JavaScript function is called, the enclave provides the interpreter with function inputs and any other state that should be available to the code about to run. 

Javascript running in the enclave can access the content of protected HTML forms via the the global variable \texttt{forms}. The forms variable contains a property for each form name. For example, with reference to the HTML code in \Cref{codePicture}, the payment form can be accessed via \texttt{forms.payment} where \texttt{payment} is the value of the attribute \texttt{name} of the \texttt{<form>} tag. Developers can implement custom input validation procedures. For example, a very simple form of validation can be checking if the credit card field contains forbidden characters such as white spaces. The JavaScript function that verifies the presence of white spaces can be implemented as shown in \Cref{formvalidation}.

\begin{lstlisting}[style=htmlcssjs, caption=\small Simple form validation, captionpos=b, label={formvalidation}]
function cardNumberHasWhiteSpaces() {
  return /\s/g.test(forms.payment.card);
}
\end{lstlisting}

\noindent\textbf{Network Communication.}
In order for protection of user data on the local machine to translate into a useful web application, there must be a mechanism for transmitting data out from the enclave without tampering by the compromised browser or OS. We provide a basic mechanism for doing this by supporting HTML forms, but web applications in general need to send back data to the server programmatically in a variety of contexts, not just when a user submits a form. To support this need, we provide support for XMLHttpRequests (as shown in \Cref{xmlhttprequest}) where requests are encrypted inside the enclave using the shared key from the attestation process before leaving the enclave. 

\begin{lstlisting}[style=htmlcssjs, caption=\small XMLHTTPRequest example, captionpos=b, label={xmlhttprequest}]
function doPay(e) {
  // input form to JS associative array
  d = toDict(forms.payment);
  
  // validate payment data
  if (validate(d)) {
      return false;
  }
  
  // prepare raw messages
  json_str = JSON.stringify(d);
  
  // create SecureXMLHttpRequest  
  var xhr = new SecureXMLHttpRequest();
  xhr.open("POST", 
    "https://pay.site.com/submit_data", 
    false); // only sync calls
  
  // use sec_json content type  
  xhr.setRequestHeader('Content-Type', 
    'application/sec_json; charset=UTF-8');
  
  // encrypt, sign, and send
  xhr.send(json_str);
  
  // seal data for possible future reuse
  storeCreditCardData(d);
}
\end{lstlisting}

The problem of defending against replay of messages over the network is not unique to the trusted hardware setting and must be handled separately by applications built on \name/. 

\noindent\textbf{Persistent Storage.}\label{storage}
\name/ provides developers with a web storage abstraction similar to the standard web storage provided by unmodified web browsers. Secure web storage can be accessed via \texttt{localStorage}, as shown in \Cref{codelisting}. 

\begin{lstlisting}[style=htmlcssjs,caption=\small Web storage, captionpos=b, label={codelisting}]
function storeCreditCardData(d){
  localStorage['holder'] = d.holder;
  localStorage['cc']     = d.card;
  localStorage['exp']    = d.expiry;
  localStorage['cvv']    = d.cvv;
}
\end{lstlisting}

When the need for persistent storage arises, \name/ encrypts the data to be stored using a \emph{sealing key} and stores it on disk (it could equivalently use existing browser storage mechanisms to hold the encrypted data). The sealing key is a feature provided by SGX to an enclave in order to store persistent data across multiple runs of the enclave. 

This approach raises two problems we must resolve. First, every instance of the same enclave shares the same sealing key, so we must ensure that different enclaves created by the same browser cannot read each others' secrets. We can prevent this problem by including the associated origin as additional authenticated data with the encrypted data to be stored. This way an enclave can find and restore data associated with the origin it connects to but, as a matter of policy, does not allow the user to access data associated with any other origin. The integrity guarantees of our trusted hardware platform ensure that our code will abide by this policy. 

The second issue is that of rollback attacks. A malicious operating system could roll back or delete data that is stored to disk, so, for applications that rely on maintaining sensitive state, the enclave must have a way to determine whether it has the most up-to-date stored data. A generic solution to this problem, such as ROTE~\cite{rote}, would suffice, but ROTE requires a distributed setting which may not be available to a user browsing the web from home. We can solve this problem by enlisting the assistance of the server to ensure protection against rollbacks, especially in situations where an enclave is connected to a server that already keeps information about the user. The idea is to keep a \emph{revision number}, one for each origin, that gets sent from the server to the enclave at the end of the attestation process and is incremented whenever changes are made to locally stored data. Since the attacker cannot change the number stored on the server or in the enclave during execution, we can detect whenever a rollback attack has been launched or stored data has been deleted by observing a mismatch between the number on the data reloaded by the enclave and the number sent by the server.

Our generic approach for storage of user secrets and network connections could easily be extended to include storage of cookies, resulting in a separate cookie store, accessible only to the enclave, that otherwise provides the same functionality available from cookies in unmodified browsers. 

\section{Security Analysis}\label{security}
In this section we give a clear enumeration of the different kinds of threats against which we expect \name/ to defend and argue that \name/ does indeed protect against these attacks. We first discuss attacks on the core features of \name/ and then move on to attacks targeted specifically at the trusted I/O path and user interface. 

\subsection{Attacks on Core Features}

\textbf{Enclave omission attack}. An attacker with full control of the sofware running on a system may manipulate the browser extension and enclave manager software to pretend to use an enclave when in fact it does not. This attack will, however, fail because of defenses built into our user interface via the keyboard and display dongles. Absent a connection to a real enclave, the trusted input lights on the keyboard and display will not light, alerting the user that entered data is unprotected.  

\textbf{Enclave misuse attack}. A more subtle attack of this form uses the enclave for some tasks but fakes it for others. For example, to circumvent the defense above, trusted input from the user could use the real enclave functionality, but trusted output on the display could be spoofed without the enclave. As such, it is necessary for each I/O device to separately defend against fake use of an enclave. The defenses described for the previous attack suffice to protect against this attack as well, but both lights are needed.

An attacker could also use the genuine trusted I/O path but attempt to omit use of the enclave when running JavaScript inside the browser. This attacker could clearly not access persistent storage, trusted network communication, or user inputs because those features require keys only available inside the enclave. On the other hand, the JavaScript to be run inside the enclave is not encrypted, so an attacker could potentially also run it outside the enclave, so long as it does not make use of any other resources or features offered by \name/. At this point, however, the JavaScript becomes entirely benign because it cannot give the attacker running it any new information or convince the user or remote server of any falsehoods because the trusted paths to all private information or trusted parties are barred. 

A last variant of this attack would omit certain ECALLs that perform necessary setup operations like initializing a form and its inputs before the user begins to enter data. Omission of these ECALLs would result in the system crashing but would not leak secrets in the process. As mentioned before, we cannot conceivably protect against a denial of service attack where the compromised OS refuses to allow any access to the system. We can only ensure that normal or abnormal use of the enclave does not leak user secrets. 

\textbf{Page tampering attack}. Failing to omit an enclave entirely or even partially, the attacker can turn to modifying the inputs given to various ecalls. In particular, the names and structure of forms and their inputs or the JavaScript to be run inside the enclave could be modified. Mounting this attack, however, would require an adversary who can break the unforgeability property of the signatures used to sign secure \texttt{<form>} and \texttt{<script>} tags. Those tags are verified with an origin-specific public key (either hard-coded in the enclave or verified with a certificate) that lies out of reach of our attacker. 

Since trusted JavaScript is the only way to access trusted user inputs from within the browser, the fact that we have separate scope for execution of trusted and untrusted JavaScript means that any attempt to directly access user secrets stored in protected inputs will necessarily be thwarted. 

\textbf{Redirection attack}. This attack resembles a straightforward phishing attempt. Instead of tampering with the operation of \name/, a browser could navigate to a malicious website designed to look legitimate in an attempt to send user secrets to an untrusted server. Here again the persistent overlay added by our display dongle prevents an attack by displaying the origin to which the enclave has connected. The strict same-origin policy within the enclave means that the origin displayed in the trusted portion of the screen is the only possible destination for network connections originating withing the enclave. While an attacker could establish a connection with a server other than the declared origin, the data sent to that server will be encrypted with a key known only to the intended origin, rendering the data useless to others. As such, the only way for an attacker to have legitimate-looking text appear there is to send user data only to legitimate destinations. 

\textbf{Storage tampering attack}. Although authenticated encryption with a sealing key tied to the enclave protects persistently stored data from tampering, an attacker can still delete or roll back the state of stored data. We detail our solution to protect against this attack in Section~\ref{storage}, where we enlist the assistance of the server to keep an up-to-date revision number for the enclave's data out of reach of the attacker. Attacks where the browser connects to a malicious site whose trusted JavaScript tries to read or modify persistent storage for other sites are prevented by our policy of strict separation between stored data associated with different origins.

\subsection{Attacks on Trusted I/O Path and UI}
We now consider attacks against the trusted I/O path to the user. Direct reading of private key presses and display outputs is prevented by encryption of data between the enclave and keyboard/display dongles, but we also consider a number of more sophisticated attacks. Since the I/O path to the user closely relates to the user interface, we discuss attacks against both the protocols and the interface together. We discuss security considerations involved in the setup of trusted I/O devices in Section~\ref{setupio}.

\textbf{Mode switching attack}. As discussed in \Cref{sec:trustedpath}, the decision to switch between trusted and untrusted modes ultimately lies with the untrusted browser because it decides when an input field receives focus or blurs or when to activate \name/ in the first place. We defend against this type of tampering with the light on the dongles and the delay when switching from trusted to untrusted modes. These defenses protect against both a standard unauthorized exit from the enclave as well as a rapid switching attack that tries to capture some key presses by quickly switching between modes. 

\textbf{Replay attack}. We defend against replay of trusted communications between the enclave and display by including a non-repeating count in every message that is always checked to make sure an old count does not repeat. An attacker could, however, eavesdrop on key presses destined for one enclave, switch to a second enclave connected with a site it controls, and replay the key presses to the second enclave in an attempt to read trusted key presses. We defend against this attack by including the name of the origin along with the count in encrypted messages, so they cannot be replayed across different enclaves. Likewise, since the keyboard and display use different keys to encrypt communications with the enclave(s), messages cannot be replayed across sources. 

\textbf{Input manipulation attack}. Attackers can attempt to make untrusted input fields appear where a user might expect trusted input fields and thereby fool users into typing trusted information in untrusted fields. Since the attacker has almost full control of what gets placed on the display, this grants considerable freedom in manipulating the display to mimic visual queues that would indicate secure fields. Fortunately, our display dongle reserves a strip at the bottom of the screen for trusted content directly from the enclave. This area informs the user what trusted input is currently focused, if any. 

An attacker could also manipulate the placement of actual trusted input fields or the labels that precede them on a page in order to confuse or mislead a user as to the purpose of each field. By using the trusted display area to show \emph{which} trusted input currently has focus, if any, we give developers the opportunity to assign inputs descriptive trusted names that will alert a user if there is a mismatch between an input's name and its stated purpose in the untrusted section of the display. 

\textbf{Timing attack}. The fact that key presses originate with the user means that the timing of presses and associated updates to content on the screen may leak information about user secrets~\cite{keytiming}. We close this timing side channel by having the keyboard send encrypted messages to the enclave at a constant rate while in trusted mode, sending null messages if the user does not press a key during a given time period and queueing key presses that appear in the same time period. A high enough frequency for this process ensures that the user experience is not disrupted by a backlog of key presses. Updates to display overlay contents also happen at a constant rate, so timing channels through key presses and display updates cannot leak information about user secrets. 

\textbf{Multi-Enclave Attacks}. 
As mentioned in Section~\ref{threatmodel}, \name/ does not aim to protect against attacks mounted by incorrect or privacy-compromising code provided by an origin that has already been authenticated. That said, we briefly discuss here some attacks that could be launched by collaboration between a malicious OS and a malicious remote origin that is trusted by \name/ (for example, in case of a maliciously issued certificate) and which tries to steal data a user meant to send to a different trusted origin. An attacker who has compromised a trusted site could always ask for data from a user directly, rendering these attacks less important in practice, but there may be some kinds of data a user would only want to reveal to one trusted origin and not others, e.g. a password for a particular site. 

First we consider an enclave-switching attack, a more involved variant of the mode-switching attack described above. In this attack, the untrusted sytem rapidly switches between different \emph{enclaves}, one connecting to a legitimate site and the other to a malicious site controlled by the attacker. \name/'s existing mode-switching delay also protects against this variant of the attack because the display always shows the origin associated with the enclave currently in use.

A more complicated attack could run one honest, uncompromised enclave concurrently with an enclave connected to an malicious origin. The uncompromised enclave would feed its overlays to the display while the compromised enclave would receive inputs from the keyboard. This may be noticed by users in the current \name/ design because anything typed would not appear on the display, but by the time a user notices this, secrets may have already been compromised. To defend against this, the keyboard and display dongles could be configured to only connect to one enclave at a time (not connecting to another enclave until the first enclave declares it has entered the end state) and to check that they have connected to the same enclave at setup by using the enclave to send each other hashes of fresh origin-specific secrets.

\section{Implementation}\label{sec:imp}

\begin{figure}
    \centering
    \includegraphics[width=.95\linewidth]{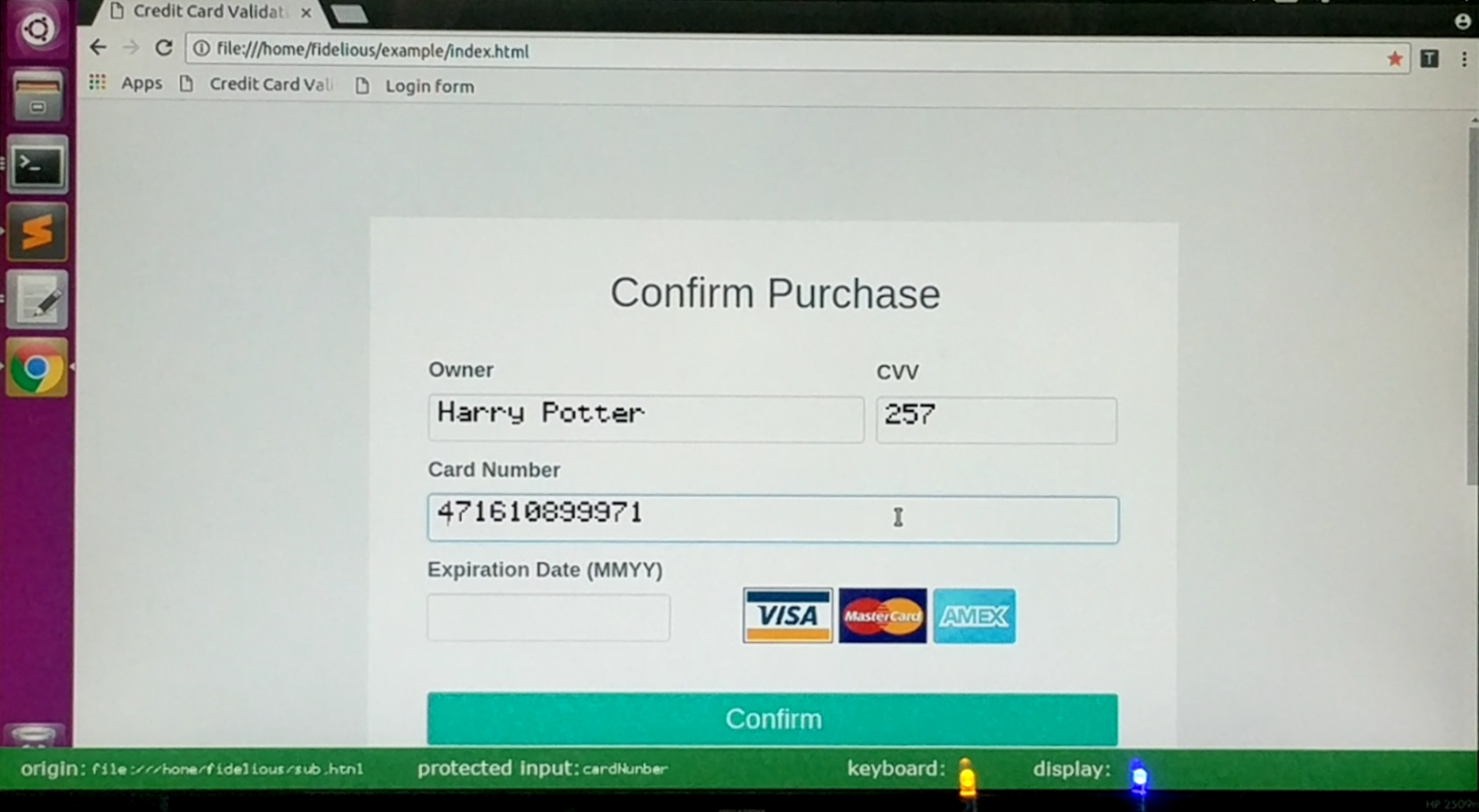}
	\includegraphics[width=.95\linewidth]{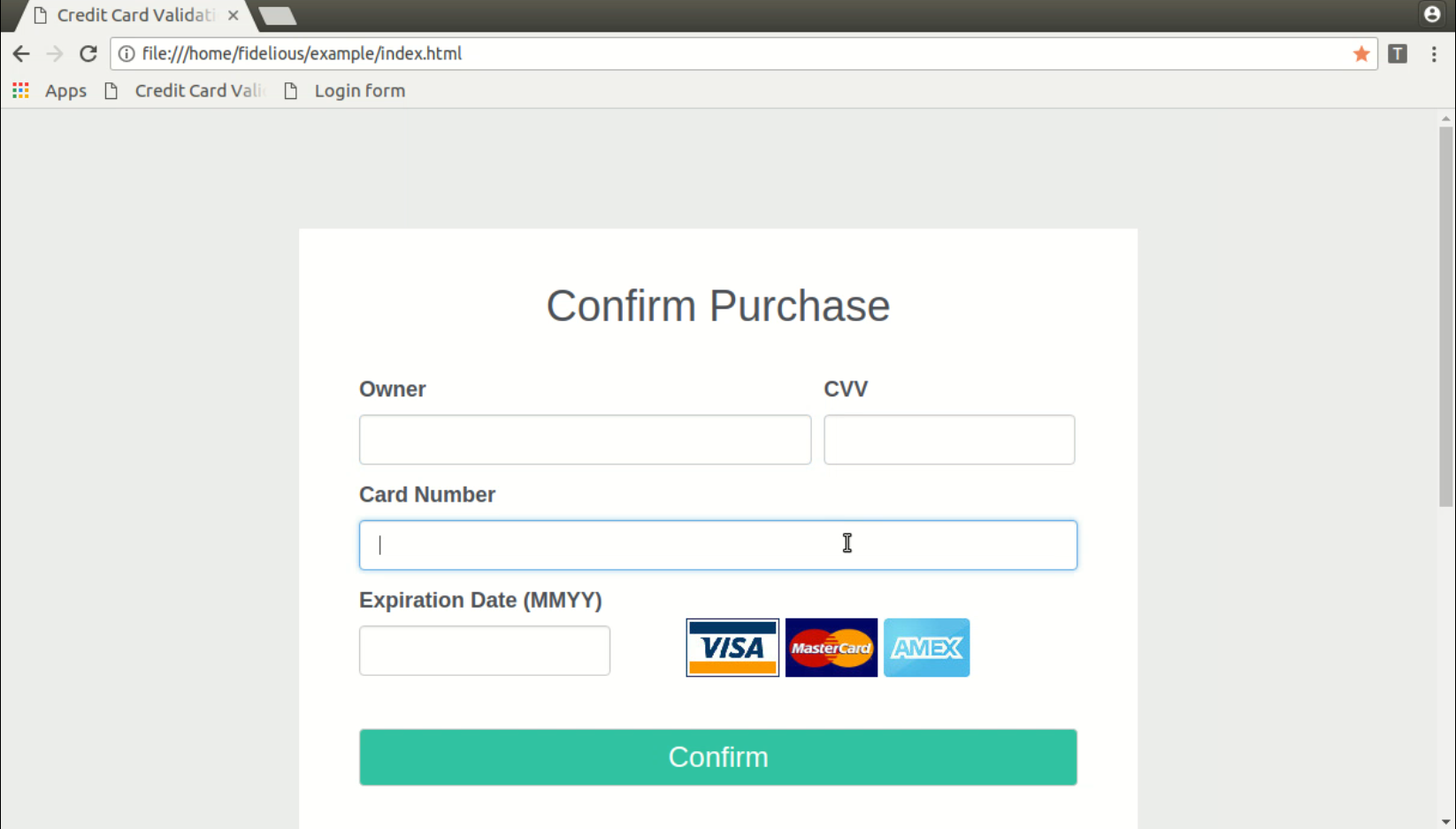}
\caption{\small Images of our \name/ prototype in use. The image above shows the view of a user, and the image below shows the view of an attacker taking a screen capture while the user enters credit card information. Since trusted overlays are decrypted and placed over the image \emph{after} leaving the compromised computer, the attacker does not see the user's data.}
\label{fig:screenshots}
\end{figure}

\begin{figure*}
	\centering
	\includegraphics[width=0.9\linewidth]{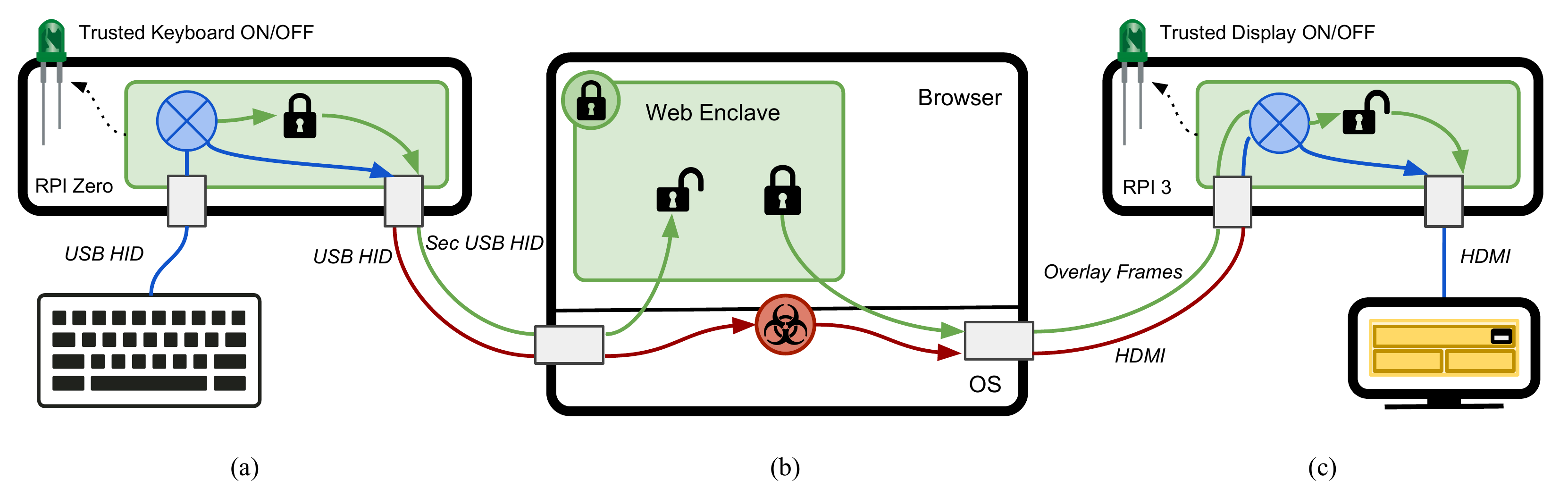}
	\caption{\small Prototype of the trusted path: (a) standard USB keyboard connected to our RPI Zero dongle to encrypt keystrokes, (b) Computer with a Fidelius-enabled browser, and (c) standard HDMI display connected to our RPI 3 dongle to overlay secure frames.}
	\label{fig:prot}
\end{figure*}

We implemented a prototype of \name/, including both the trusted path described in Sections \ref{UIsection} and \ref{sec:trustedpath} and the Web Enclave features described in \Cref{features}\footnote{Our open source implementation of \name/, the instructions to build the dongles and accompanying sample code are available at \url{https://github.com/SabaEskandarian/Fidelius}.}. Our prototype is fully functional but does not include the trusted setup stage between the enclave and devices, which we carry out manually. \Cref{fig:screenshots} shows screenshots of our prototype in use, and \Cref{fig:prot} gives an overview of its physical construction. 

Since \name/ requires few changes on the server side and our evaluation therefore focuses on browser overhead, we do not implement a server modified to run \name/. This would consist mainly of having the server verify a remote attestation and decrypt messages from the web enclave.

\subsection{Trusted Path}\label{imppath}

\begin{figure}
\centering
	\includegraphics[width=0.9\linewidth]{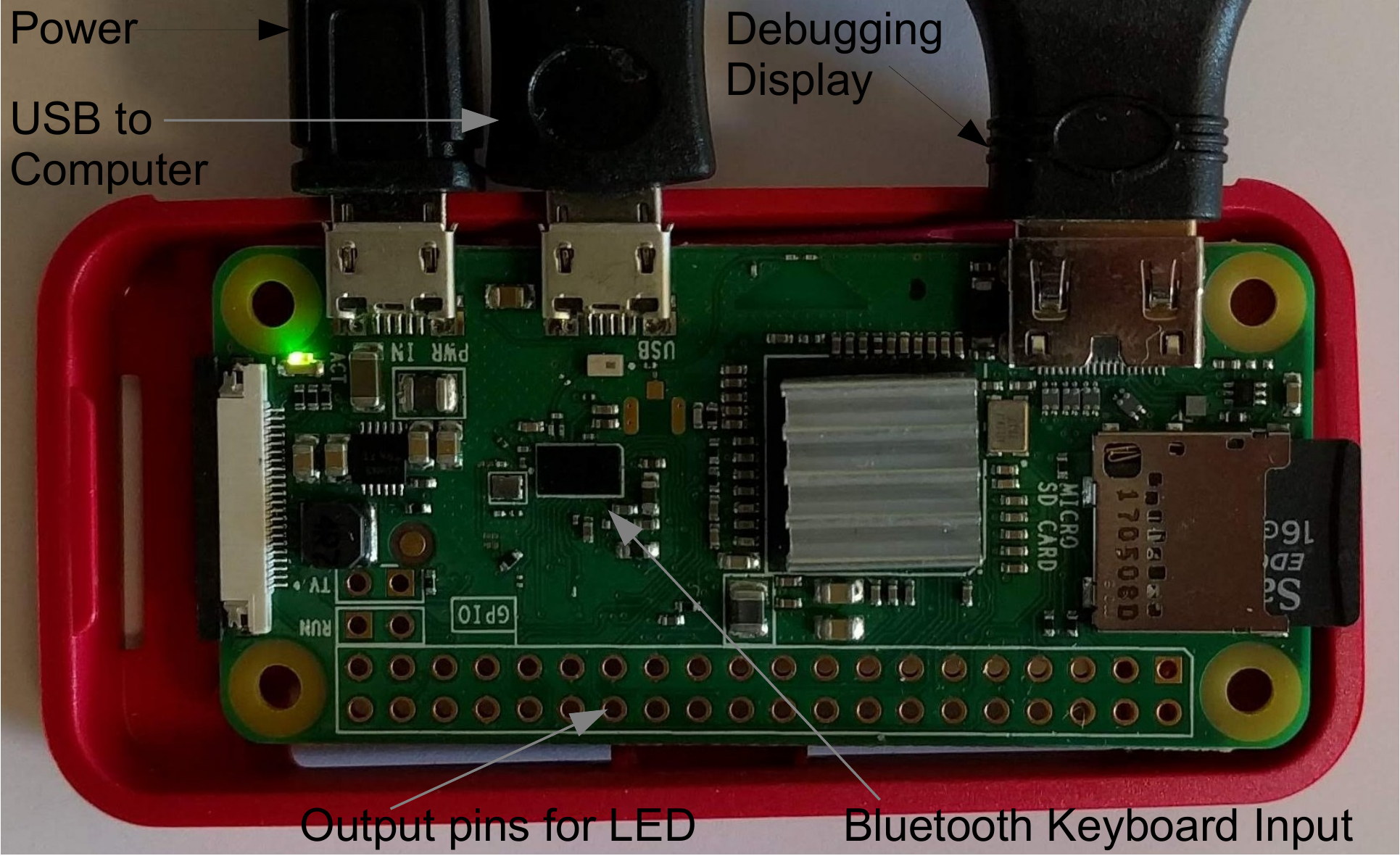}
	\caption{\small Trusted keyboard dongle built from Raspberry Pi Zero. In untrusted mode, the dongle forwards key presses from the keyboard to the computer. In trusted mode, the dongle sends a constant stream of encrypted values to the enclave. The values correspond to key presses if there has been any input or null values otherwise.}
	\label{fig:rpi0}
\end{figure}
\begin{figure}
\centering
	\includegraphics[width=0.9\linewidth]{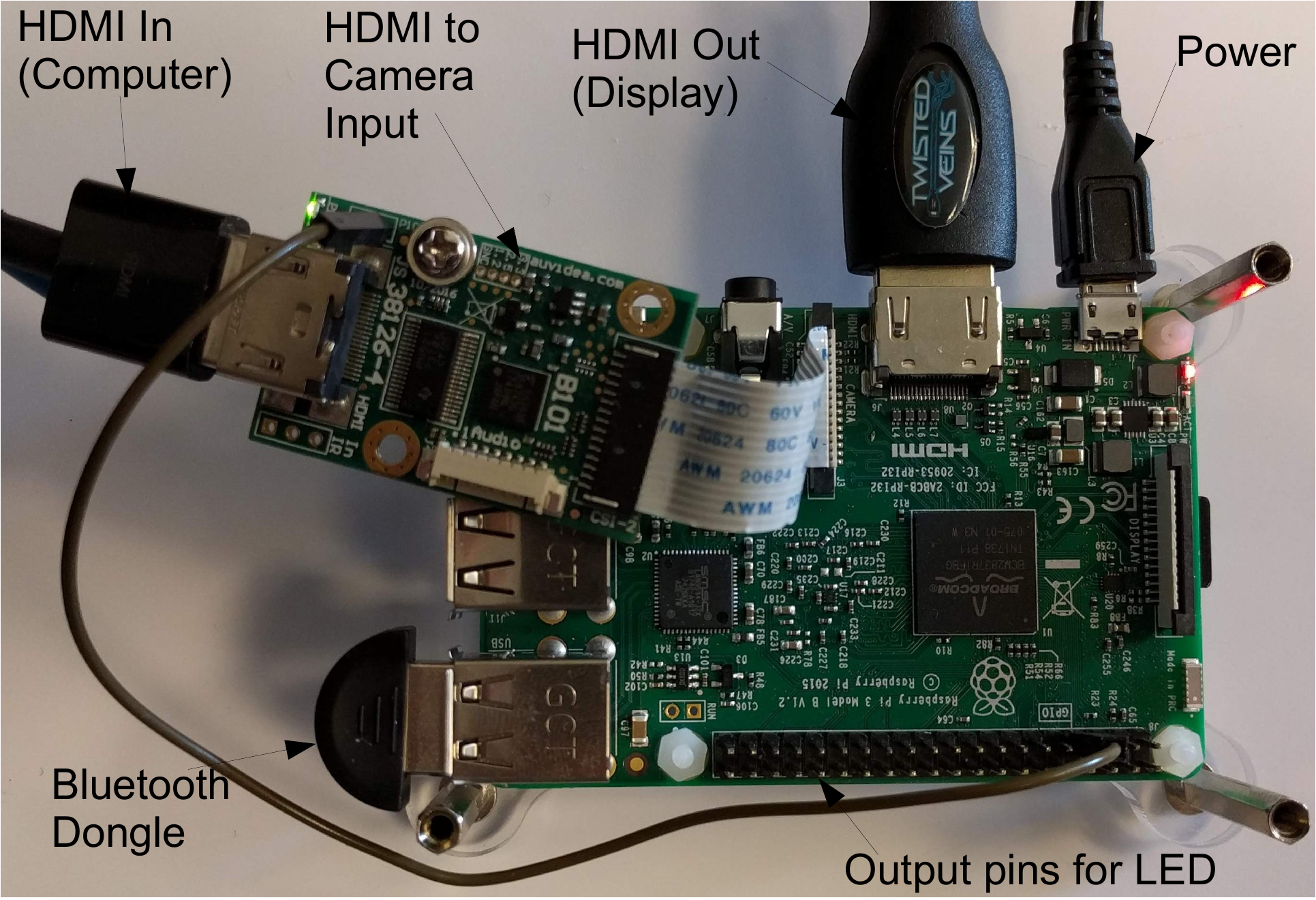}
	\caption{\small Trusted display dongle built from Raspberry Pi 3. Frames arrive on the RPI3 over HDMI in, which connects through a board that treats the frames to be displayed as camera inputs. Overlays are transmitted over Bluetooth and decrypted on the RPI3. The combined frame and overlay go to the display through the HDMI out cable.}
	\label{fig:rpi3}
\end{figure}

Our prototype runs on an Intel Nuc with a 2.90 GHz Core i5-6260U Processor and 32 GB of RAM running Ubuntu 16.04.1 and SGX SDK version 2.1.2. We produced dongles to place between the Nuc and an off-the-shelf keyboard and display using a Raspberry Pi Zero with a 1 GHz single core Broadcom BCM2835 processor and  512 MB of RAM  running Raspbian GNU/Linux 9 (stretch) for the keyboard and a Raspberry Pi 3 with a 1.2 GHZ quad-core ARM Cortex A53 processor and 1GB RAM running Raspbian GNU/Linux 9 (stretch) at a display resolution of 1280x720. Figures~\ref{fig:rpi0} and~\ref{fig:rpi3} show our input and output dongle devices. 

The Raspberry Pi Zero simulated two input devices to the Nuc, one standard keyboard and one secure keyboard, with only one device active at any time based on the state of the application being run. 
The RPI 3 uses a B101 rev. 4 HMDI to CSI-2 bridge and the Picamera Python library~\cite{picamera} to treat the HDMI output from the Nuc as a camera input on which it overlays trusted content before rendering to the real display. Trusted content is sent over a separate bluetooth channel. The bluetooth channel exists as a matter of convenience for implementation, as HDMI does allow for sending auxiliary data, but we were unable to programmatically access this channel through existing drivers.

When an encrypted overlay packet reaches the RPI3 display device from the Nuc, it is first decrypted and decoded from a flat black and white encoding used to transfer data back to a full RBG color representation. Next, the image is transferred from the decryption/decoding program to the rendering code, which places it on the screen. We introduce a refresh delay between sending frames to give the Picamera library adequate time to render each frame before receiving the next one.

Although we have built a working \name/ prototype, a number of improvements could make for a more powerful and complete product. These changes include miniaturization of dongle hardware, faster transfer protocols, e.g. USB 3.0 instead of Bluetooth, and custom drivers to reduce latency between the dongles and the keyboard/display. We leave the engineering task of optimizing \name/ to future work. 

\subsection{Browser and Web Enclave}

On the Intel Nuc device, \name/ is implemented as a Chrome browser extension running on Chrome version 67.0.3396 communicating with a native program via Chrome's Native Messaging API\footnote{See \url{https://developer.chrome.com/apps/nativeMessaging}} for web enclave management. The extension activates on page load and checks whether the page contains components that need to be protected, e.g., secure HTML forms and JavaScript. If it does, it communicates with the native program to initiate the web enclave and perform remote attestation with the server. Once this process completes, the user can interact with secure components on the page, and secure JavaScript code can be run in the enclave. Since the page setup process occurs independently of the page loading in the browser, only the secure components of a page are delayed by the attestation process -- non-secure elements of a page have no loading penalty as a result of running \name/.

The majority of the work of enclave management is handled by the native code. For symmetric encryption of forms, bitmaps, and keystrokes we use AES-GCM encryption and for signing forms we use ECDSA signatures.  JavaScript inside the enclave is run on a version of the \texttt{tiny-js}\cite{tinyjs} interpreter that we ported to run inside the enclave.

\section{Evaluation}\label{sec:eval}
\begin{figure*}
\begin{minipage}{.32\textwidth}
\small
\begin{tikzpicture}
		\begin{axis}[
				width=6cm,
				ybar,
				y label style={at={(axis description cs:.15,.5)},anchor=south},
				title={Display Response Latency},
				ylabel={Latency (ms)},
				extra y ticks=201.82,
				extra y tick labels={},
				extra y tick style={
						ymajorgrids=true,
						ytick style={
								/pgfplots/major tick length=0pt,
							},
						grid style={
								red,
								dashed,
								/pgfplots/on layer=axis foreground,
							},
					},
				ticklabel style = {align=center, font=\scriptsize},
				symbolic x coords={iPhone \\6s,Galaxy \\S7,Fidelius,HTC Rezound,Kindle Oasis 2},
				xtick=data, 
				x tick label style={text width=1.1cm},
				ytick={0,200,400,600},
				ymin=0,ymax=600,
				bar width=20.
			]
			\addplot+[text=black] coordinates {(iPhone \\6s,70) (Galaxy \\S7,120) (Fidelius,201.82) (HTC Rezound,240) (Kindle Oasis 2,570)};

		\end{axis}
\end{tikzpicture}
\end{minipage}
\begin{minipage}{.32\textwidth}
\small
\begin{tikzpicture}
	\begin{axis}[
	width=6cm,
	y label style={at={(axis description cs:.19,.5)},anchor=south},
	title={Additional Page Load Time},
	ytick={0,20,40,60},
    ymin = 0,
    ymax = 70,
    ticklabel style = {align=center, font=\scriptsize},
    ylabel ={Latency (ms)},
    xlabel={Number of Trusted Components}
	]
	\addplot[mark=star,color=red] table [scatter, x=num, y=Scripts, col sep=comma] {pageloaddata.csv};
	\addplot[mark=*,color=blue] table [scatter, x=num, y=Forms, col sep=comma] {pageloaddata.csv};
	\legend{Scripts,Forms}
	\end{axis}
\end{tikzpicture}
\end{minipage}
\begin{minipage}{.32\textwidth}
\small
\begin{tikzpicture}
		\begin{axis}[
				title={\name/ Display Pipeline Costs},
				width=6cm,
				ybar,
				y label style={at={(axis description cs:.18,.5)},anchor=south},
				ylabel={Latency (ms)},
				ticklabel style = {align=center, font=\scriptsize},
				symbolic x coords={Refresh\\,Decrypt,Decode,Transfer,Render},
				xtick=data,
				ymin=0,
				bar width=20,
			]
			\addplot+[text=black] coordinates {(Refresh\\,75.555) (Decrypt,2.1795) (Decode,6.9225) (Transfer,21.7145) (Render,91.446)};
		\end{axis}
\end{tikzpicture}\vspace{.01cm}
\end{minipage}
\begin{minipage}{.32\textwidth}
		\caption{\small \name/ key press to display latency compared with the screen response time on various commercial devices. }
	\label{compcom}
\end{minipage}\hspace{.2cm}
\begin{minipage}{.32\textwidth}
	\caption{\small \name/'s impact on page load time as the number of trusted components varies. Adding components does not significantly affect load time.}
	\label{pageload}
\end{minipage}\hspace{.2cm}
\begin{minipage}{.32\textwidth}
		\caption{\small Breakdown of display costs by component. Render/refresh delays are an artifact of our hardware and could be dramatically reduced.}
	\label{components}
\end{minipage}
\end{figure*}

We evaluate \name/ in order to determine whether the overheads introduced by the trusted I/O path and web enclave are acceptable for common use cases and find that \name/ outperforms display latency on some recent commercial devices by as much as 2.8$\times$ and prior work by 13.3$\times$. Moreover, communication between the browser and enclave introduces a delay of less than 40ms to page load time for a login page. We also identify which components of the system contribute the most overhead, how they could be improved for a production deployment, and how performance scales for larger and more complex trusted page components.

\noindent\textbf{TCB Size.}
The trusted code base for \name/ consists of 8,450 lines of C++ code, of which about 3200 are libraries for handling form rendering and another 3800 are our enclave port of \texttt{tiny-js}. This does not include native code running outside the enclave or in the browser extension because our security guarantees hold even if an attacker could compromise those untrusted components of the system. It also excludes dongle code which runs on the Raspberry Pi devices and not the computer running the web browser. Compared to the 18,800,000 lines of the Chrome project\footnote{\url{https://www.openhub.net/p/chrome/analyses/latest/languages_summary}}, \name/ supports many of the important functionalities one may wish to secure in a web browser while exposing an attack surface orders of magnitude smaller than a naive port of a browser into a trusted execution environment.

\noindent\textbf{Comparison to Commercial Devices.}
For a standard login form with username and password fields, \name/'s key press to display latency is 201.8 ms. We exclude the time it takes to transfer the encrypted key press from the keyboard to the enclave over USB 2.0 (480 Mbps) and the encrypted bitmap from the enclave to the display over bluetooth (3 Mbps) from these figures. This is a reasonable omission because the size of the data being transferred is small compared to the transfer speed of these two protocols. \Cref{compcom} compares the latency between a key press and display update in \name/ to measurements of the display latency on several commercial mobile devices~\cite{inputlag}. Although not competitive with high-performance devices, \name/ performs comparably or even faster than some popular commercial devices, running 2.8$\times$ faster than the latency on the most recent Kindle. Fidelius's efficiency arises from leaving the majority of a page unmodified and only using encrypted overlays for trusted components.

\noindent\textbf{Comparison to Prior Work.}
We also compared \name/ to Bumpy~\cite{bumpy}, which provides a trusted input functionality but no corresponding display. For this comparison, we compared Bumpy to \name/'s trusted path without the display component, which accounts for the vast majority of the latency. Bumpy's source code is not available, so we compare to the reported performance values measured on an HP dc5750 with an AMD Athlon64 X2 Processor at 2.2 GHz and a Broadcom v1.2 TPM. \name/ outperforms Bumpy's reported performance by 13$\times$, running with a latency of 10.59ms compared to Bumpy's 141ms. We believe this more than compensates for differences in the computing power used to evaluate the two systems. Although SGX-USB~\cite{sgxusb}, whose source code is also unavailable, was developed on more recent hardware, we cannot compare directly to their reported performance results because they report generic USB data transfer rates into an enclave whereas we care about the latency of reading and processing key presses. 

\noindent\textbf{Page Load Overhead.}
\Cref{pageload} shows the page load overhead incurred by Fidelius, not including remote attestation. \name/'s overhead  includes the time for the browser to inform the enclave of secure components and for the enclave to verify signatures on them, totaling 35.3ms. We do not report time for remote attestation, which depends on the latency to the attestation service. Fortunately, waiting for the attestation server to respond can occur in parallel with other page load operations because notifying the enclave of the existence of trusted components and verifying signatures do not involve sensitive user data. Moreover, attestation time is independent of page content, so our measurements fully capture \name/'s page load time increase as trusted components are added. As seen in \Cref{pageload}, adding components does not significantly increase page load time.

\begin{figure}
	\centering
	\small
    \setlength{\tabcolsep}{4pt}
    \setlength\extrarowheight{2pt}
    \begin{tabular}{l | r r | r | r r}
		\toprule
        \multicolumn{1}{c|}{Field size(s)} & \multicolumn{1}{c}{W} & \multicolumn{1}{c|}{H} & \multicolumn{1}{c|}{W$\times$H px} & Time (ms) & Incr. (ms) \\
		\midrule
		1 Small                           & 171                   & 50                    & 8,550         & 195.83    & -     \\
		1 Medium                          & 342                   & 50                    & 17,100        & 199.20    & 3.38  \\
		1 Large                           & 683                   & 50                    & 34,150        & 209.65    & 10.45\\
		\hline
		1 Extra large                     & 911                   & 50                    & 45,550        & 214.74    & -     \\
		2 Extra large                     & 911                   & 100                   & 91,100        & 227.02    & 12.28  \\
		\bottomrule
    \end{tabular}
	\caption{\small Key press to display latency when rendering forms. Widths are fractions of the most popular screen width ($w=1366$px): S=$\frac{1}{8}w$, M=$\frac{1}{4}w$, L=$\frac{1}{2}w$, XL=$\frac{2}{3}w$. Increments calculated from the previous row.}
	\label{overlaysize}
\end{figure}

\noindent\textbf{Performance Factors.}\label{componentanalysis}
\Cref{components} shows the cost of various components of our trusted display pipeline, described in \Cref{imppath}, which makes up almost all of \name/'s performance overhead. The two most expensive operations that take place on the display are rendering the overlay using the Picamera module and the refresh delay we introduce in order to allow the Picamera module to process frames without forming a queue of undisplayed frames. The Picamera module and associated hardware on the Raspberry Pi 3 is not optimized to add a dynamic overlay to the camera feed. A better approach would involve directly manipulating the data from the Nuc computer's HDMI output instead of using it to simulate a camera and placing overlays on top of the camera feed. This could easily be achieved in a production deployment of \name/ and would dramatically reduce display latency.

We also considered how performance varies as the size of the trusted components on a page increase. \Cref{overlaysize} shows that latency increases linearly with the size of the trusted component. This happens because as the size of the overlay increases, it takes longer to decrypt, decode and transfer the overlays. Taking steps to optimize the display pipeline would further mitigate latency increase. However, even under our current implementation, for two full-page width input fields (See the two extra large input field experiments in \Cref{overlaysize}), \name/ has a display latency of only 227ms. Also, a tenfold increase in pixels  (from one small field to two extra large fields) results in only a 31ms latency increase.

\section{Discussion and Extensions}\label{discussion}

\name/ opens the door to a new class of secure web applications supported by the widespread availability of hardware enclaves in modern computers. The fundamental problems solved by \name/ -- reliably establishing a path from I/O devices to an enclave residing in an otherwise untrusted system and of protecting web applications without moving large portions of a browser into an enclave -- have applications well beyond the login and payment examples described thus far. 

\name/'s techniques and architecture can also support more complex applications such as online tax filing or even web-based instant messaging. The trusted I/O path has applications beyond the web as well and could be adapted to secure logins or desktop applications that use enclaves for their core functionality but require interaction with a local user on the machine. We anticipate that \name/'s I/O approach will be very useful, as hardware enclaves are most widely available on consumer desktop and laptop computers. 

We close with a discussion of possible extensions that could broaden the applicability of our architecture or would be important considerations in a widespread deployment. 

\noindent\textbf{Usability of Trusted Devices.} 
We have implemented \name/ with a user and developer interface that provides users with the necessary tools to interpret their interaction with \name/ properly and avoid UI-based attacks. However, our interface represents only one possible design for interaction between users and the core \name/ functionality. A great deal of work has studied the effectiveness of security indicators such as our indicator lights~\cite{WI05,emperor}. Other possible designs may, for example, use secure attention sequences or separate trusted buttons to initiate communication with trusted components. Future work could explore this space to determine what approach works best for this application in practice. 

\noindent\textbf{Event Loop.}
\name/ leaves the JavaScript event loop outside the enclave to optimize the tradeoff between TCB size and functionality. A number of additional applications could be enabled by moving the event loop into an enclave, especially if there is a way to accomplish this more efficiently than with a direct port that executes the loop as-is in trusted hardware. 

\noindent\textbf{HTML Rendering.}
In order to render HTML forms, we implemented a custom library that, given a description of a form and its inputs, produces a bitmap that represents the form. In order to extend support to other HTML tags, we need to integrate a more versatile rendering engine into our web enclaves. Existing libraries such as Nuklear~\cite{nuklear} provide a solid first step in this direction. 




\noindent\textbf{Root Certificate Store.}
Our current implementation of the web enclave uses a limited number of public keys. To scale to supporting any web site, the web enclave needs to have a root certificate store inside the enclave.

\noindent\textbf{Mobile Devices.}
We have described \name/ in the setting of a desktop device, but much of users' interaction with the web today takes place on mobile devices. While much of the \name/ architecture could apply equally well in an enclave-enabled mobile setting, a trusted path system for phones and tablets will necessarily look very different from the keyboard and display dongles used by \name/. Android's recent protected confirmation system~\cite{android-protected} represents a promising first step in this direction. 

\section{Related Work}\label{sec:related}

\noindent \textbf{NGSCB}. In 2003 Microsoft announces the Palladium
effort, later renamed NGSCB~\cite{NGSCB}.  In that design, attestation 
is provided by a TPM chip and enclave isolation is provided by 
hardware memory curtaining. The project was scaled back in 2005
presumably due to the difficulty of adapting applications to the architecture.
In contrast, as we explained, web sites can take advantage of \name/
by simply adding an HTML attribute to web fields and forms that it wants
to protect. 

\noindent \textbf{SGX and the Web}. TrustJS~\cite{trustjs} explores the potential for running JavaScript inside an enclave, demonstrating that running trusted JavaScript on the client-side can expedite form input validation. SecureWorker~\cite{secureworker} provides the developer abstraction of a web worker while executing the worker's JavaScript inside an enclave. Our work uses the ability to run JavaScript in an enclave as a building block to enable privacy for user inputs in web applications. JITGuard~\cite{JITGUard} uses SGX to protect against vulnerabilities in Firefox's JIT compiler.  

\noindent \textbf{Unmodified Applications on SGX}. A handful of works aim to allow execution of unmodified applications inside an SGX enclave. Haven runs whole applications inside an enclave~\cite{haven}, while SGXKernel~\cite{sgxkernel}, Graphene~\cite{graphene}, and Panoply~\cite{panoply} provide lower level primitives on which applications can be built. Scone~\cite{scone} secures linux containers by running them inside an enclave. Flicker~\cite{flicker} and TrustVisor~\cite{trustvisor} use older hardware to provide features similar to SGX on which general applications can be built, albeit with weaker performance due to the older and more limited hardware features on which they build. We focus on directly solving the problem of hiding user inputs in an untrusted browser without using generic solutions in order to minimize TCB and avoid the potential pitfalls of porting a monolithic browser into a trusted environment. 

\noindent \textbf{SGX Attacks and Defenses}. A number of side channel attacks on SGX have been shown to take advantage of, among other things, memory access patterns~\cite{cacheattacks,controlledchannelattacks,BMD+17}, asynchronous execution~\cite{asyncshock}, branch prediction~\cite{branchshadowing}, speculative execution~\cite{sgxpectre,foreshadow}, and even SGX's own security guarantees~\cite{SWG+17} to compromise data privacy. There do, however, exist many defenses that have been shown to evade these side channels, often generically, without a great deal of overhead~\cite{raccoon,SCNS16,sgxshield,tsgx,rote,moat}. Even more promising, researchers have proposed a series of other architectures~\cite{sanctum,ghostrider,phantom} which defend against weaknesses in SGX \emph{by design} and are therefore invulnerable to broad classes of attacks. As our work is compatible with generic defenses and concerns itself primarily with higher level functionalities built over enclaves, we do not consider side channels in the presentation of \name/. 

\noindent \textbf{Protection Against Compromised Browsers}. A number of software-based solutions for protection against compromised browsers offer tradeoffs between security, performance, and TCB size. Shadowcrypt~\cite{shadowcrypt} uses a Shadow DOM to allow encrypted input/output for web applications, but is vulnerable to some attacks~\cite{crackingshadowcrypt}. Terra~\cite{terra} uses VMs to allow applications with differing security requirements to run together on the same hardware. Tahoma~\cite{tahoma}, IBOS~\cite{ibos}, and Proxos~\cite{proxos} integrate support for browsers as OS-level features, allowing smaller TCBs and stronger isolation/security guarantees than in a general-purpose OS. Cloud terminal~\cite{cloudterminal} evades the problem of local malware and protects against attackers by only running a lightweight \emph{secure thin terminal} locally and outsourcing the majority of computation to a remote server. 

\noindent \textbf{Trusted I/O Path}. While many works study how to use a hypervisor to build a trusted path to users (e.g.~\cite{BT17,ZGNM12,YGZ15,tip,nitpicker}), little work has been done in the trusted hardware setting. SGXIO~\cite{sgxio} provides a hybrid solution that combines SGX with hypervisor techniques to allow a trusted I/O path with unmodified devices. In contrast, our work relies \emph{only} on hardware assumptions with no need for a hypervisor, but does require modified keyboard and display devices. Intel has alluded to an internal tool used to provide a trusted display from SGX~\cite{sgxappwhitepaper,sgxvidconf}, but no details, source code, or applications are available for public use. SGX-USB~\cite{sgxusb} allows for generic I/O but does not solve the problem of mixing trusted and untrusted content in a user interface as we do in both our keyboard and display. ProximiTEE~\cite{proximitee} bootstraps a similar generic trusted I/O path off of a modified attestation procedure with new safeguards over standard SGX attestation. 

Bumpy~\cite{bumpy} (and its predecessor BitE~\cite{bite}) use the trusted execution environment provided by Flicker~\cite{flicker} to provide a secure input functionality similar to ours. Aside from the larger web architecture which we build over our trusted I/O features, we go beyond these works by 1) enabling interactivity with the trusted input via the trusted display (Bumpy does not display characters the user types) and 2) closing timing side channels on user input (an improvement we also offer over SGX-USB).

\section{Conclusion}
We have presented \name/, a new architecture for protecting user secrets from malicious operating systems while interacting with web applications. \name/ protects form inputs, JavaScript execution, network connections, and local storage from malware in a fully compromised browser. It also features the first publicly available system for a trusted I/O path between a user and a hardware enclave without assumptions about hypervisor security. Our open source implementation of \name/, accompanying sample code, and a video demo are available at \url{https://github.com/SabaEskandarian/Fidelius}.

\section*{Acknowledgment}

We thank Amit Sahai and Keith Winstein for helpful conversations about
this work. This work was supported by NSF, DARPA, a grant from ONR,
the Simons Foundation, a Google faculty fellowship, and the German Federal Ministry of Education and Research (BMBF) through funding for the CISPA-Stanford Center for Cybersecurity (FKZ: 13N1S0762).

\bibliographystyle{IEEEtran}
\footnotesize
\bibliography{sgxweb} 

\end{document}